\documentclass[twocolumn]{aastex63}
 
\accepted{by ApJ Letters}

\shorttitle{Heavy elements in Leo ring}
\shortauthors{Corbelli et al.}

\begin{document}

\title{Heavy elements unveil the non primordial origin of the giant HI ring in Leo}

\correspondingauthor{Edvige Corbelli}
\email{edvige@arcetri.inaf.it}

\author[0000-0002-5788-2628]{Edvige Corbelli}
\affiliation{INAF-Osservatorio di Arcetri, Largo E. Fermi 5, 50125 Firenze, Italy}

\author[0000-0002-5281-1417]{Giovanni Cresci}
\affiliation{INAF-Osservatorio di Arcetri, Largo E. Fermi 5, 50125 Firenze, Italy}
 
\author[ 0000-0002-4803-2381]{Filippo Mannucci}
\affiliation{INAF-Osservatorio di Arcetri, Largo E. Fermi 5, 50125 Firenze, Italy}

\author[0000-0002-8528-7340]{David Thilker}
\affiliation{Department of Physics and Astronomy, The Johns Hopkins University, Baltimore, MD, USA}

\author[0000-0001-8349-3055]{Giacomo Venturi}
\affiliation{Instituto de Astrof\'{i}sica, Facultad de F\'{i}sica, Pontificia Universidad Cat\'{o}lica de Chile, Casilla 306, Santiago 22, Chile}
\affiliation{INAF-Osservatorio di Arcetri, Largo E. Fermi 5, 50125 Firenze, Italy}

\begin{abstract}

The origin and fate of the most extended extragalactic neutral cloud known in the Local Universe, the Leo ring, is still debated 38 years after its discovery. Its existence is alternatively attributed to leftover primordial gas with some  low level of  metal pollution versus enriched gas stripped during a galaxy-galaxy encounter. 
Taking advantage of MUSE (Multi Unit Spectroscopic Explorer) operating at the VLT, we performed optical integral field spectroscopy  of 3 HI clumps in the Leo ring where ultraviolet continuum emission has been found. We detected, for the first time, ionized hydrogen in the ring and identify 4 nebular regions powered by massive stars. These nebulae show several  metal  lines ([OIII], [NII], [SII])  which allowed  reliable measures of metallicities, found to be close to or above the solar value  (0.8$\le Z/Z_\odot \le$1.4). Given the faintness of the diffuse stellar  counterparts, less than 3$\%$ of the observed heavy elements could have been produced locally in the main body of the ring and not much more than 15$\%$  in the HI clump towards M\,96. This inference, and the chemical homogeneity among the regions, convincingly demonstrates that the gas in the ring is not primordial, but has been pre-enriched in a galaxy disk, then later removed and shaped by tidal forces and it is forming a sparse population of stars. 

\end{abstract}


\section{Introduction} \label{sec:intro}

The serendipitous discovery of an optically dark HI cloud in the M\,96 galaxy group \citep{1983ApJ...273L...1S}, part of the Leo I group,  has since then triggered a lot of discussion on the origin and survival of the most  massive and extended intergalactic neutral cloud known in the local Universe ($D\le 20$~Mpc). With an extension of about 200~kpc and an HI mass $M_{HI} \simeq 2\times 10^9$~ M$_\odot$, the cloud has a ring-like shape  orbiting the galaxies M\,105 and NGC\,3384 \citep{1985ApJ...288L..33S}, and it is known also as the Leo ring.
As opposed to tidal streams, the main body of the Leo ring is isolated, more distant than 3 optical radii from any luminous galaxy. The ring is much  larger than any known ring galaxy \citep{2008MNRAS.386L..38G}. The collisional ring of NGC\,5291 ($D\simeq 50$~Mpc) \citep{1979MNRAS.188..285L,2007A&A...467...93B}, of similar extent, is  vigorously forming stars, as many other collisional rings. The Leo ring is much more quiescent and for many years since its discovery has been detected only via HI emission.  Lacking a  pervasive optical counterpart  \citep{1985AJ.....90..450P,1985MNRAS.213..111K,2014ApJ...791...38W} it has been proposed as a  candidate primordial cloud \citep{1989AJ.....97..666S, 2003ApJ...591..185S} dating to the time of the Leo I group formation.  

The bulk of the HI  gas in the ring is on the south and west side, especially between M\,96 (to the south) and NGC\,3384/M\,105 (at the ring center, see Figure~1). Intermediate resolution VLA  maps  of this region with an effective beam of 45\arcsec\  revealed the presence of gas clumps \citep{1986AJ.....91...13S},  some of which appear as distinct virialized entities and have masses  up to 3.5$\times 10^7$~M$_\odot$. The position angle of the clump major axes and their velocity field suggest  some internal rotation with a possible disk-like geometry and gas densities similar to those of the interstellar medium.  Distinct cloudlets are found in the extension pointing south, towards M\,96. Detection of GALEX UV-continuum light in the direction of a few HI clumps  of the ring, suggested star formation activity between 0.1 and 1~Gyr ago \citep{2009Natur.457..990T}. However, most of the gas mass is not forming massive stars today since there has been no confirmed diffuse H$\alpha$ emission   \citep{1986ApJ...309L...9R,1995ApJ...450L..45D}  or CO detection from a pervasive population of giant molecular complexes \citep{1989AJ.....97..666S}.

A low level of metal enrichment, inferred from GALEX-UV and optical colors,  favoured  the primordial origin hypothesis. This was supported  a few years later by the detection of  weak metal absorption lines in the spectra of 3 background QSOs,  2 of which have sightlines close or within  low HI column density contours  of the ring \citep{2014ApJ...790...64R}.  The low metallicity, estimated between 2$\%$ - 16$\%$ solar for Si/H, C/H and N/H, has however large uncertainties due to ionisation corrections. Confusion with emission from the Milky Way in the QSO's spectra does not allow to measure HI column densities along the sightlines. This is inferred from large scale HI gas maps, and gas substructures on small scales can alter the estimated abundance ratios.

\begin{figure*} 
\hspace*{-1.5 cm}
\includegraphics [width=14.8 cm,angle=-90]{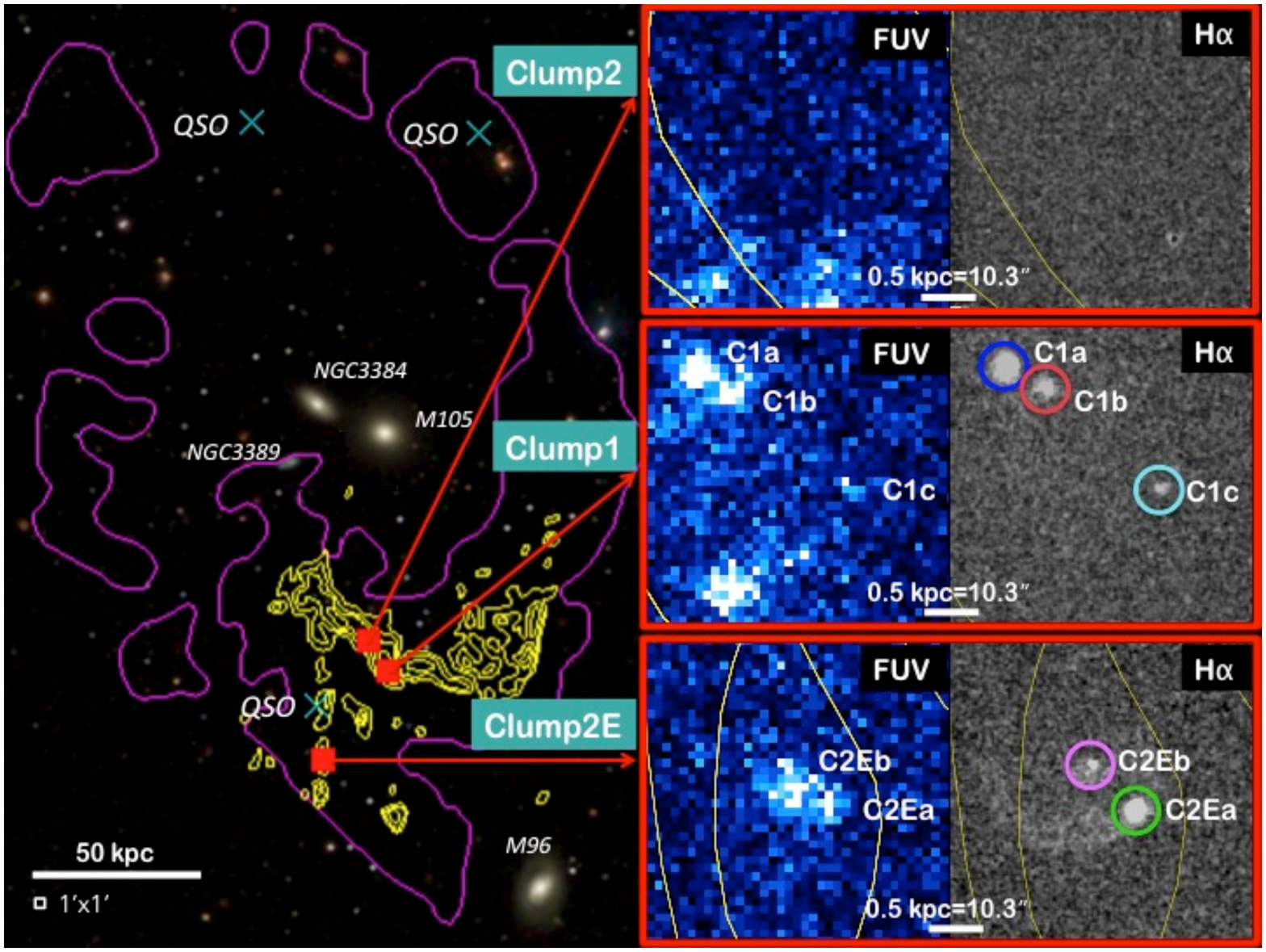}
\caption{In {\it left panel} the HI contours of the Leo ring are overlayed to the optical image of the M96 group (SDSS color image).  In magenta the Arecibo contour at N$_{HI}=2\times 10^{18}$~cm$^{-2}$, in yellow the VLA HI contours of the southern part  of the ring \citet{1986AJ.....91...13S}. Crosses indicates background QSOs \citep{2014ApJ...790...64R},  red squares the HI clumps observed with MUSE: Clump1(C1), Clump2(C2) and Clump2E(C2E).  The 1~arcmin$^2$ angular size of the MUSE field is shown in the bottom left corner. In the {\it right panels} the MUSE H$\alpha$  images  of Clump1  and of Clump2E  show the 5  nebular regions detected. The corresponding GALEX-FUV continuum emission is displayed to the left of the H$\alpha$ images. The southernmost FUV  source in the field of Clump1 is a background galaxy. 
}
\label{fields}
\end{figure*}

The Leo ring has also been considered as a possible  product of a gas-sweeping head-on collision \citep{1985ApJ...288..535R},  involving group members such as NGC\,3384 and M\,96  \citep{2010ApJ...717L.143M} or a low surface brightness galaxy colliding with the group \citep{2005MNRAS.357L..21B}. A tentative detection of dust emission at 8~$\mu$m in one HI clump (Clump1)  \citep{2009AJ....138..452B} also supports the pre-enrichment scenario. A direct and reliable measurement of high metallicity gas associated to the very weak stellar counterpart can give the conclusive signature of a ring made of pre-enriched  gas. 

In this Letter we present the first detection of nebular regions in the Leo ring. In Section~2 we describe integral field optical spectroscopy of 3 fields in the ring and estimate metal abundances from emission lines in star forming regions.  The local metal production and the implications for the origin of the Leo ring are discussed in the last Section. 
In a companion paper \citep{submitted}(hereafter Paper II) we analyse star formation  and the stellar population in and around the detected nebulae using GALEX and HST images. 

\section{The discovery of nebular regions and their chemical abundances}

We assume a distance to  the Leo ring of 10~Mpc, as for M\,96 and M\,105. This implies that an angular separation of 1\arcsec\  corresponds to a spatial scale of 48.5~pc.

\subsection{The data}

Between December 2019 and March 2020  we have observed three 1$\times$1~arcmin$^2$ regions in the Leo ring using the integral field spectrograph  MUSE (Multi Unit Spectroscopic Explorer) mounted on the ESO Very Large Telescope (VLT). The locations of MUSE fields are shown in red in the left panel of Figure~\ref{fields} overlaid on the SDSS optical image of the M\,96 group and on the VLA  HI contours of the ring. The fields have been centered at 3 HI peak locations,  Clump1, Clump2  and Clump2E, two in the  main body  of the ring and one in the filament connecting the ring to M\,96. They cover completely the  ultraviolet-bright  regions  of Clump1 and Clump2E listed by \citet{2009Natur.457..990T}. The southernmost side of the UV emission in Clump2 is at the border of the MUSE field.

The final cube for each region is the result of two observing blocks, one totalling 960~s and the other 1920~s. The observing blocks are a combination of two and four 480~s exposures, respectively, which were  rotated and dithered  from each other in order to provide a uniform coverage of the field and to limit systematics. Dedicated offset sky exposure of 100~s each were acquired every two object exposures. The reduction of the raw data was performed with  the  ESO MUSE pipeline \citep{2020A&A...641A..28W}, which includes the  standard procedures of bias subtraction, flat fielding, wavelength calibration, flux calibration, sky subtraction and the final cube reconstruction by the spatial arrangement of the individual slits of the image slicers. For Clump1 we did not employ the dedicated sky observations for the sky subtraction, since these were giving strong sky residuals, especially around the H$\alpha$ line. We thus extracted the sky spectrum to be subtracted from within the science cube, by selecting the portions of the FoV free of source emission. This allowed to remove the problematic sky residuals because the sky spectrum obtained from within the science FoV is simultaneous with the science spectra. The final dataset comprises 3 data cubes, one per clump,  covering a FoV slightly  larger than 1~arcmin$^2$. Each spectrum spans the wavelength range 4600 - 9350~\AA, with a typical spectral resolution between 1770 at 4800~\AA\  and 3590 at 9300~\AA.  The spatial resolution given by the seeing is of the order of 1\arcsec. 

\subsection{HII regions in the ring}

We analyse spectral data at the  observed spectral and spatial resolution searching for H$\alpha$ emission at the  velocities of the HI gas in the ring i.e. between 860 and 1060~km~s$^{-1}$.  We detect  hydrogen  and  some  collisionally excited metal lines  in three distinct regions of Clump1 (C1a, C1b, C1c)  and in two regions of Clump2E (C2Ea, C2Eb).  Figure~\ref{fields} shows the GALEX-FUV continuum and  the H$\alpha$ emission in the  three MUSE fields. The FUV emission in Clump2E seems more extended than the HII region in H$\alpha$ and  suggests a non coeval population or the presence of some low mass stellar cluster lacking massive stars. No nebular lines are detected in the field covering Clump2. This clump is the reddest of the three clumps observed,  having the largest values of  UV and optical  colours \citep{2009Natur.457..990T,2014ApJ...791...38W}.   

\begin{table*}
\movetableright=-2.cm
\caption{HII region coordinates, chemical abundance and extinction. Extinction corrected total H$\alpha$ luminosities are computed using circular apertures with radius R$_{ap}^{max}$.} 
\centering  
 \begin{tabular}{cccccccccccc}           
\hline\hline 
Source& RA & DEC &$ V_{hel}$ & 12+log(O/H)   & $A_V$ & $Z/Z_\odot$ &   $R^{max}_{ap}$ & $A^{Rmax}_{H\alpha}$ & log $L_{H\alpha}$  \\   
             & & &  km~s$^{-1}$ &   & mag &    & arcsec & mag & erg~s$^{-1}$    \\   
 \hline\hline 
C1a& 10:47:47.93 & 12:11:31.9 &     994$\pm$2 &    8.59$^{+0.04}_ {-0.04}$       &1.02$^{+0.60}_{-0.60}$            & 0.79 &  5.0    & 0.40  & 36.62  \\
C1b& 10:47:47.44 & 12:11:27.6 &   1003$\pm$3 &    8.63$^{+0.28}_{-0.04}$        & 0.06$^{+1.59}_{-0.06}$           & 0.87 &  3.0    & ....      & 35.94  \\
C2Ea& 10:48:13.52 & 12:02:24.3&   940$\pm$3 &    8.84$^{+0.01}_{-0.01}$         & 0.47$^{+0.31}_{-0.35}$           & 1.41 &  3.4    & 0.61  & 36.91   \\
C2Eb& 10:48:14.08& 12:02:32.5&   937$\pm$21&   8.82$^{+0.09}_{-0.11}$          &  ....                                                & 1.35  &  3.0   & ....      & 35.85   \\
 \hline\hline 
\end{tabular}
\label{metals}
 \end{table*}

\begin{table*}
\movetableright=-2.cm
\caption{Integrated emission for Gaussian fits to nebular lines with R$_{ap}$=1.2\arcsec. Upper limits are 3$\sigma$ values, flux units are 10$^{-17}$~erg~s$^{-1}$cm$^{-2}$.} 
\centering                                       
\begin{tabular}{c c c c c c c c c }           
\hline\hline 
 Source & H$\beta$   &   [OIII]5007  & [NII]6548  & H$\alpha$  &  [NII]6583 & [SII]6716/  & [SII]6731 & FWHM$_{b,r}$[\AA] \\   
                             \hline\hline 
C1a&     1.89$\pm$0.37      &  1.53$\pm$0.38    & $<0.46$           &  7.89 $\pm$0.29  &  1.39$\pm$0.25   &  0.86$\pm$ 0.21   & 0.54$\pm$ 0.21   & 2.5,2.5\\
C1b&     1.00$\pm$0.37      &  $<0.76$               & $<0.49$            &   3.17$\pm$0.23    & 0.82$\pm$0.31    &  $ <0.40$                & $<0.40$              &  1.9,2.2\\
C2Ea&     7.97$\pm0.41$   &  1.34$\pm$0.37  & 3.88$\pm0.31$ & 26.57$\pm$0.35    & 11.39$\pm$0.36  &  2.71$\pm$0.33    &  1.81$\pm$ 0.36  & 2.8,2.4\\
C2Eb&     $<0.79$               &  $<0.79   $            & $<0.76 $            &  2.25$\pm$0.35     &  1.01 $\pm$0.32   &   $<0.34$               &  $<0.34$               & ...,2.1\\
 \hline\hline 
\end{tabular}
\label{lines}
\end{table*}

The four regions listed in Table~1 are  HII regions associated with recent star formation events according to their line ratios \citep{2003MNRAS.346.1055K,2012ApJ...758..133S}  and to the underlying stellar population (see Paper II).  The data relative to the faintest nebula detected, C1c, is presented and discussed in Paper II because  emission line ratios  and [OIII]5007 luminosity  are consistent with the object being a Planetary Nebula whose metallicity is unconstrained  due to undetected lines.  We give in Table~1 the central coordinates of the HII regions and  the mean recession velocities of  identified optical lines. These are consistent with the 21-cm line  velocities of the HI gas \citep{{1986AJ.....91...13S}}.  We fit Gaussian profiles to emission lines whose peaks are well above  3$\sigma$ in circular apertures with radius 1.2\arcsec, comparable to the seeing. With these apertures  we sample more than one third of the region total H$\alpha$ luminosity and achieve good signal-to-noise (S/N$>$2.5) for integrated Gaussian line fits to all detected lines. The integrated line fluxes are shown in Table~2.  We require a uniform Gaussian line width in the red or in the blue part of the spectrum since lines are unresolved. The resulting FWHM  are shown in Table~2. Upper limits in Table~2 are 3$\sigma$ values for non-detected lines, inferred using the rms of the spectra at the expected wavelength  and a typical full spectral extent of the line. For the brightest HII regions we detected strong metal lines, such as [O\,\textsc{iii}]5007, [N\,\textsc{ii}]6583, [S\,\textsc{ii}]6716,6731 which can be used to compute reliable metallicities.

\subsection{Chemical abundances}

\begin{figure} 
\vspace*{-0.7cm}
\includegraphics [width=8.4cm]{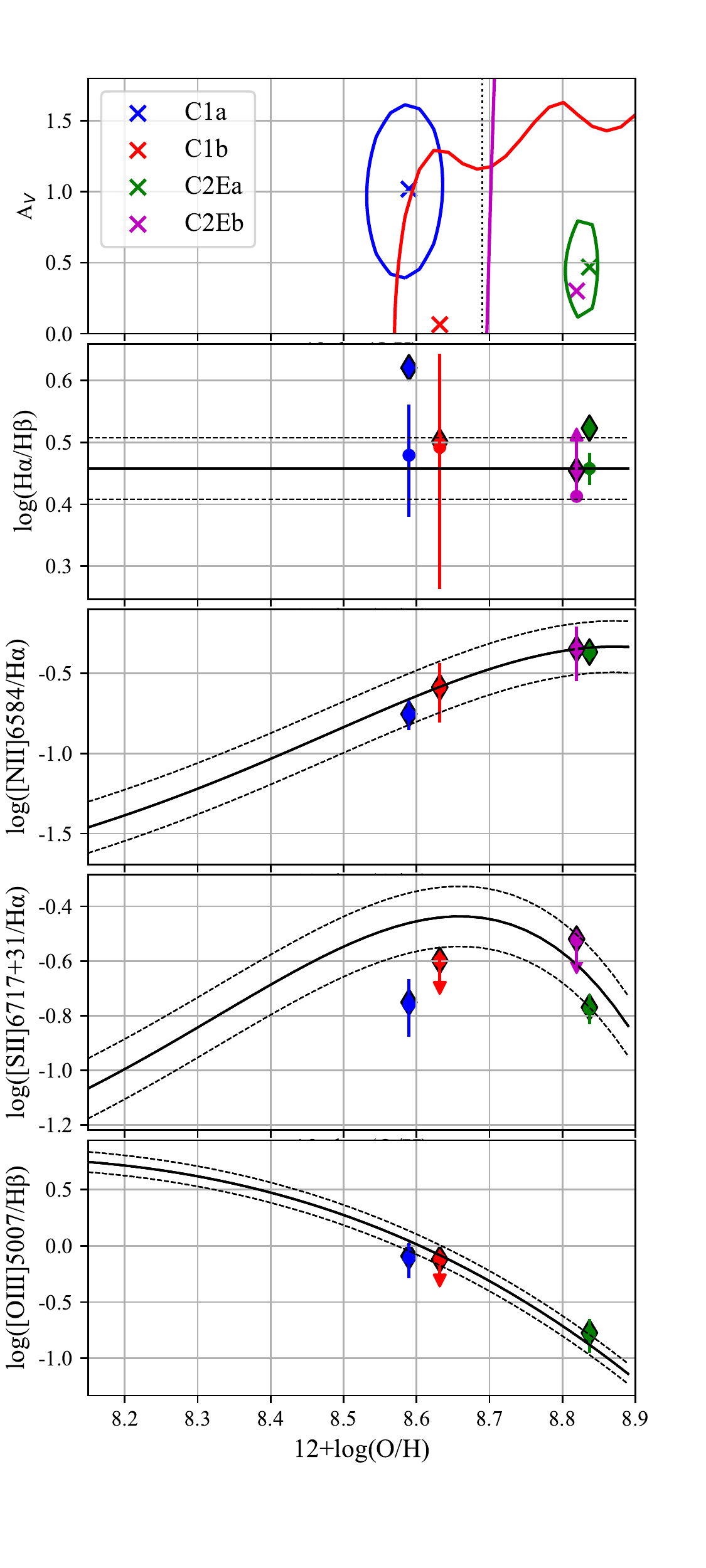}
\caption{Dust visual extinction  and gas-phase metallicity for each HII region, color-coded as in the legend. In the top panel we show the best fitting values (cross) and 1-$\sigma$ confidence levels of $A_V$; the dotted line shows the solar  metallicity (12+log(O/H)=8.69).
The four bottom panels refer to relevant strong-line ratios. Diamonds show the observed values of the line ratio plotted at the best-fitting value of metallicity. The H$\alpha$/H$\beta$ ratio is computed for Case B recombination with uncertainties due to the unknown temperature, and circles showing  extinction-corrected values. The solid curves in the lower three panels trace the calibrations from \cite{2020MNRAS.491..944C}, with the relative uncertainties. 
\label{fig_metals}
}
\end{figure}

For the  four HII regions  in Table~1 we compute the gas-phase metal abundances using the strong-line calibration in \cite{2020MNRAS.491..944C}.  All the available emission lines and the upper limits to the undetected lines are used to measure metallicity and dust extinction in a two-parameter minimization routine which also estimates the uncertainties on these two parameters. The resulting metal abundances  are displayed in Figure~\ref{fig_metals} and in  Table~1.  In Figure~\ref{fig_metals} we show the 1-$\sigma$ confidence levels in the oxygen abundance-visual extinction plane and the best fitting values of chemical abundances  along the calibration curves for the strong line ratios. Line ratios for all the HII regions  are well-reproduced by close to solar metallicities and moderate visual dust extinctions.  For C2Eb extinction cannot be constrained. Metallicities in Clump1 are  slightly below solar, those in Clump~2E are above solar. The HII regions  in Clump~1 have lower SII/H$\alpha$ line ratios than predicted by  \cite{2020MNRAS.491..944C}. This is also found in outer disk HII regions \citep{1996MNRAS.280..720V} and it is likely due to a high ionisation parameter driven by a low density interstellar medium \citep{2013ApJS..208...10D}.
The mass fraction of metals with respect to solar, $Z/Z_\odot$,  ranges between 0.79 and 1.41 (assuming solar  distribution of heavy elements and $Z_\odot$=0.0142 \citep{2009ARA&A..47..481A}).

Using wide apertures,  as listed in column~(8) of Table~1 and chosen to include most of H$\alpha$ emission with no overlap, we derive the HII region total H$\alpha$ luminosities, $L_{H\alpha}$. These are given in column~(10) already corrected for extinction when this can be estimated from the Balmer decrement in these apertures (column~(9)). Luminosities are high enough to require the presence of very massive and young stars, especially for C1a and C2Ea. The local production rate of ionizing photons by hot stars might be higher than what can be inferred using $L_{H\alpha}$ if some photons leak out or are directly absorbed by dust in the nebula.  

The HII regions in the [OIII]/H$\beta$ versus [NII]/H$\alpha$ plane, known as the BPT diagram \citep{1981PASP...93....5B},   are consistent  with data from  young HII regions in galaxies \citep{2003MNRAS.346.1055K,2012ApJ...758..133S} and their metallicities are in agreement with those predicted by photoionisation models of HII regions  \citep{2013ApJS..208...10D,2017ApJ...840...44B}.
Line ratios observed in Clump1 are also consistent with the distribution of the recent evolution models of HII regions in gas clouds \citep{2020MNRAS.496..339P}, available only for solar metallicity. These  predict an age of about 5~Myrs for C1a. 
Line ratios for C2Ea instead fall outside the area where solar metallicity HII regions are found, in agreement with the higher than solar metallicity we infer for this clump.  A very young age is recovered in Paper II for this HII region through a multiwavelength analysis.

 \section{In situ metal enrichment and the ring origin}

We compute the maximum mass  fraction of metals which could conceivably be produced in situ, $f_Z^{max}$, given the observed metal abundances, $Z_{obs}$, and the limiting blue magnitudes of the Leo ring, $\mu_B$. For this extreme local enrichment scenario we assume that all stars have formed in the ring and use the instantaneous burst or continuous star formation models of  Starburst99  \citep{1999ApJS..123....3L}  in addition to population synthesis models of  \citet{2003MNRAS.344.1000B} for an initial burst with an exponential  decay  ($\tau=1$~Gyr). At each time step we compute the $B-V$ color and the maximum stellar surface mass density which corresponds to the limiting values of $\mu_B$. This stellar density gives the maximum mass fraction of metals produced locally. In order to maximise  the local metal production we consider a closed box model with no inflows or outflows  for which a simple equation relates the stellar yields to the increase in metallicity  since star formation has switched on  \citep{1972ApJ...173...25S}:

\begin{equation}
f_Z^{max}={Z\over Z_{obs}}= {y_Z\over Z_{obs}}\  \hbox{ln}({\Sigma_{g0}\over \Sigma_g}) = {y_Z\over Z_{obs}}\  \hbox{ln}(1+{\Sigma_*\over \Sigma_g}) 
\end{equation}

where $Z$ and  $\Sigma_*$ are  the  abundance of metals by mass and the stellar mass surface density produced in situ.  The  gas mass surface density at the present time and at the time of the Leo ring formation are $\Sigma_{g}$ and $\Sigma_{g0}$ respectively. The total net yields $y_Z$ refers to the mass of all heavy elements produced and injected into the interstellar medium by a stellar population to the rate of  mass locked up into low mass stars and stellar remnants. There are several factors that can affects the yields: the upper end of the Initial Mass Function (hereafter IMF),  massive star evolution and ejecta models,  metallicity. Since the pioneer work of \citet{1972ApJ...173...25S} several papers have analysed these dependencies \citep[e.g.][]{1992A&A...264..105M,2002A&A...390..561M,2010A&A...522A..32R,2016MNRAS.455.4183V}.  
Following the results of \citet{2010A&A...522A..32R,2016MNRAS.455.4183V} we consider negligible the metallicity dependence on the yields and consider the Chabrier IMF i.e. an IMF with a Salpeter slope from 1~$M_\odot$ up to its high mass end at 100~$M_\odot$ and a Chabrier-lognormal slope from 0.1 to 1~$M_\odot$ \citep{1955ApJ...121..161S,2003PASP..115..763C}. This  IMF has a total yield $y_Z$=0.06, the highest amongst commonly considered IMF  \citep{2016MNRAS.455.4183V}. 

\begin{figure} 
\includegraphics [width=8.5 cm]{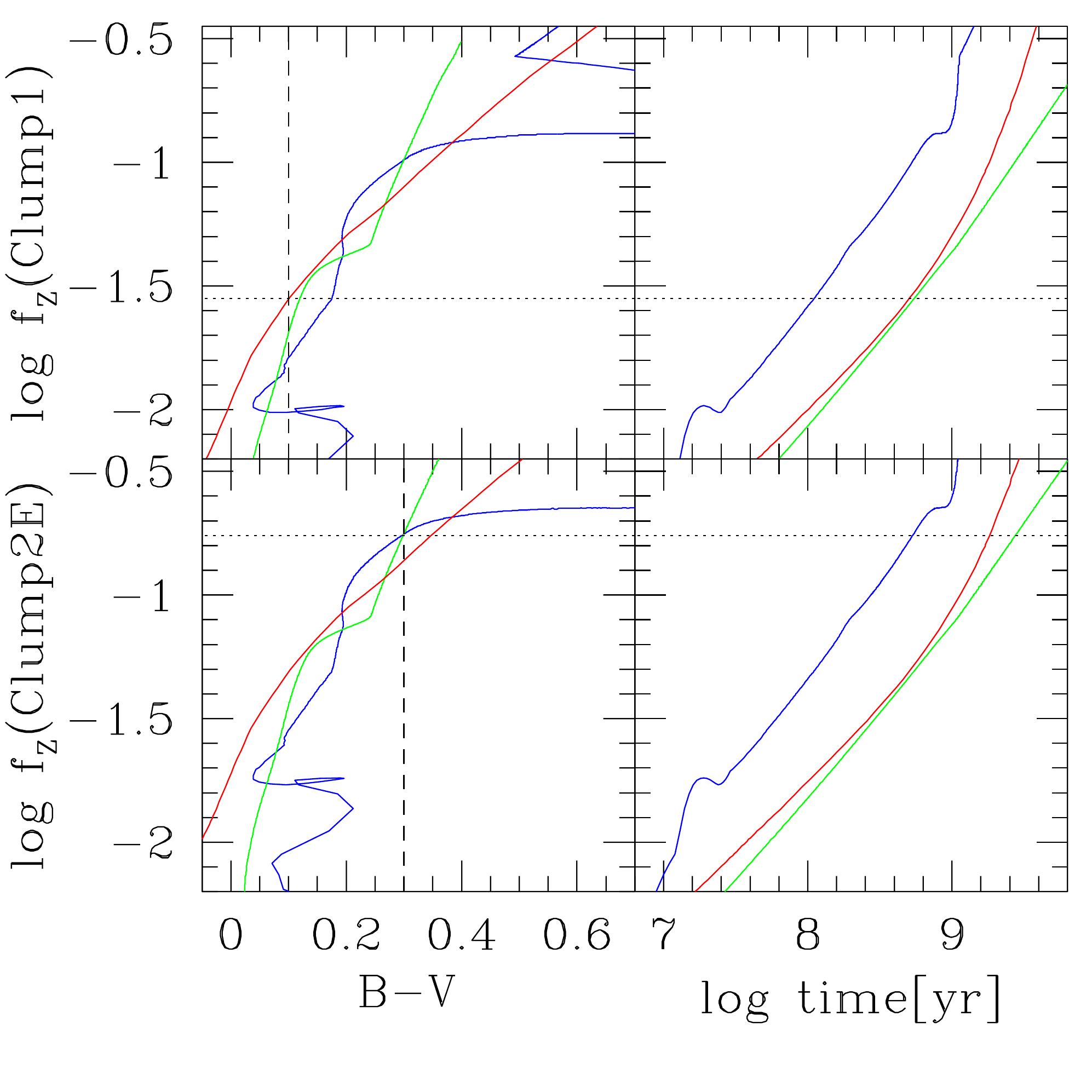}
\caption{The mass fraction of metals $f_Z$ produced in situ for an instantaneous burst (blue lines), a burst with exponential decay (red lines), and a continuous star formation model (green lines)  as a function of optical colors (left panels) and time (right panels). Each model for both Clump1 (upper panels) and Clump2E (lower panels) is normalised  as to produce an apparent magnitude $\mu_B=29$ at any time after star formation switches on. The dashed lines indicate the maximum $B-V$ optical color of the clumps, and the dotted line  $f_Z^{max}$, the highest value of $f_Z$ compatible with $B-V$. }
\label{fz}
\end{figure}

To maximize the associated fraction of metals produced locally,  $f_Z^{max}$, we consider zero extinction and the best fitted metallicities for C1a and C2Ea minus 3 times their dispersion, i.e. $Z_{obs}$=0.6 and 1.32~$Z_\odot$ for Clump1 and Clump2E respectively.  A very large fraction of the HI rich ring area corresponding to the VLA coverage of \citet{1986AJ.....91...13S} has been surveyed deeply in the optical B band  \citep{1985AJ.....90..450P,2014ApJ...791...38W}.  For a very diffuse pervasive population throughout the Leo ring the survey results give $\mu_B\ge 30$~mag~arcsec$^{-2}$. For optical emission in less extended regions as the MUSE fields, or equivalently at the VLA HI map spatial resolution \citep{1986AJ.....91...13S}, and following the results of \citet{2018ApJ...863L...7M}  we can use the more conservative upper limit $\mu_B\ge 29$~mag~arcsec$^{-2}$.  Given the  optical colors $B-V$=0.0$\pm0.1$ for Clump1 and $B-V$=0.1$\pm 0.2$~mag for Clump2E \citep{2014ApJ...791...38W} we consider  $B-V\le$0.1 and  $B-V\le$0.3  for Clump1 and Clump2E respectively. The average HI+He gas surface density over a circular area with 45\arcsec\ radius is $\Sigma_{g}$=3.1 and 0.8~$M_\odot$~pc$^{-2}$  in Clump1 and Clump2E  respectively.  We compute from the models the stellar mass surface density  corresponding to $\mu_B=29$~mag~arcsec$^{-2}$ at each time, and $f_Z$ with the above values of $\Sigma_{g}$,  $Z_{obs}$ and $y_Z$,  using equation (1). The value of $f_Z^{max}$ will be $f_Z$ at the maximum value of $B-V$ for each clump. In Figure~\ref{fz} we show $f_Z$ for the three models as a function of time and of $B-V$. The dashed line indicates the limiting value of $B-V$. A dotted line has been placed at the value of $f_Z^{max} $ i.e. where the limiting colors intersect the models which produces the highest mass of metals. 

For Clump1 a starburst 500~Myrs ago  that slowly decays with time gives the highest possible  local metal production  with $f_Z^{max}$ = 3$\%$ and $\Sigma_*$=0.01~$M_\odot$~pc$^{-2}$. For Clump2E both an instantaneous burst 500~Myrs ago or a continuous star formation since 2~Gyr ago gives the maximum value of $f_Z^{max}$ = 17$\%$ with $\Sigma_*$=0.04~$M_\odot$~pc$^{-2}$.  We conclude that the fraction of metals  produced locally is   too small to be compatible with a scenario of a primordial metal poor ring enriched in situ. The ring must have  formed out of  metal rich gas, with chemical abundances above 0.5~$Z_\odot$,  mostly polluted while residing in a galaxy and then dispersed into space. 

We underlines that all models predicts  a small  fraction of metals produced in situ and that the ones that maximise $f_Z$ are not necessarely the best fitted models to the underlying stellar population. These will be examined in Paper II. The apparent discrepancy between our results and the lower abundances  inferred by QSO's absorption lines can be resolved if  hydrogen column densities along  sightlines to nearby QSOs are lower than those used in the analysis of \citet{2014ApJ...790...64R} and estimated from HI emission averaged over a large beam.  The most discrepant abundance with respect to the nearly solar abundances we infer for the ring  is for carbon toward the southernmost QSO: -1.7$\le$ [C/H]/[C/H]$_\odot \le -1.1$. If future measures of the HI column density towards the QSO's sightline confirm the low metal abundances, these can be used to investigate  chemical inhomogeneities due to a mix of  metal rich gas  with local intragroup metal poor gas in the ring outskirts. 

We summarise that our finding has confirmed spectroscopically the association between  stellar complexes detected in the UV-continuum  and the high column density gas \citep{2009Natur.457..990T}. The detected H$\alpha$ emission implies a sporadic presence of a much younger and massive stellar population  then estimated previously (see Paper II for more details). For the first time we have detected gaseous nebulae in the ring with chemical abundances close to or above solar which conflict with the primordial origin hypothesis of the Leo ring. A scenario of pre-enrichment,  where the gas has been polluted by the metals produced in a galaxy  and subsequently  tidally stripped and placed in ring-like shape, is in agreement with the data presented in this Letter. This picture is  dynamically consistent with numerical simulations  showing the possible collisional origin of the Leo ring \citep{2005MNRAS.357L..21B,2010ApJ...717L.143M} and with chemical abundances in  nearby galaxies  possibly involved in the encounter such as M\,96 and  NGC\,3384  \citep{1993ApJ...411..137O,2007MNRAS.377..759S}.   

\begin{acknowledgements} 
EC acknowledges support from PRIN MIUR 2017 -$20173ML3WW_00$ and Mainstream-GasDustpedia. GV acknowledges support from ANID programs FONDECYT Postdoctorado 3200802 and Basal-CATA AFB-170002.

\end{acknowledgements}

\end{document}